# Odd Magneto-Optical Linear Dichroism in a Magnetophotonic Crystal

Tatiana Mikhailova [1,*], Daria Ignatyeva [1,2,3], Sergey Lyashko [1], Vladimir Berzhansky [1] and Vladimir Belotelov [1,2,3]

[1] Institute of Physics and Technology, V.I. Vernadsky Crimean Federal University, Simferopol 295007, Crimea
[2] Faculty of Physics, Lomonosov Moscow State University, Moscow 119991, Russia
[3] Russian Quantum Center, Moscow 121205, Russia
* Correspondence: taciamikh@gmail.com

**Abstract:** The phenomena of magneto-optical polarization rotation and circular magnetic dichroism are well known in the Faraday configuration. We present another effect, an odd magneto-optical linear dichroism, arising in nanostructures with polarization-dependent mode *Q*-factors. It reveals itself as the magneto-optical modulation of light intensity for the two opposite magnetization directions in the Faraday configuration. The effect was demonstrated on a magnetophotonic crystal with a cavity mode, the polarization-dependent *Q*-factor of which is due to oblique incidence. For a polarization angle of 60° (or 120°) and an angle of incidence around 60°, the magneto-optical intensity modulation maximizes and reaches 6%.

## 1. Introduction

Magneto-optical effects associated with a change in the intensity of reflected or transmitted light, called intensity effects, are widely used in integrated optics devices to create optical non-reciprocal elements [1–3], spin wave detectors [4], chemical sensors and biosensors [5–12], magnetometers [13–17] and visualization of magnetic structures [18,19].

There are several mechanisms leading to the magneto-optical intensity modulation in different configurations of the external magnetic field and light. The most known and studied is the intensity transversal magneto-optical Kerr effect (TMOKE) discovered in 1896 [20,21]. The effect is defined by the relative change of the reflectance or transmittance for the two opposite directions of an in-plane magnetization. The effect occurs in smooth magnetic films due to the influence of the external magnetic field on boundary conditions, and TMOKE typical values are less than 0.1%. TMOKE is only observed with p-polarized oblique incident light. TMOKE is significantly enhanced up to tens of percent in the materials combining gyrotropic, plasmonic and nanostructuring features due to the excitation of surface or guided modes with non-reciprocal dispersion shift [22–31]. A large TMOKE was demonstrated in all-dielectric nanostructures [32,33]. Recently, a similar effect in transverse configuration was observed for s-polarized light, too, due to the peculiarities of polarization distribution inside a nanostructure [34–37,16].

In longitudinal and polar configurations, two magneto-optical effects linear in magnetization arise in reflected light for an arbitrary linear (p + s) polarization at oblique incidence. These effects originate from the difference in the Fresnel coefficients for p- and s-polarized light and the consequent difference in the intensity of the reflected arbitrary linear polarized light under clockwise and counterclockwise magneto-optical rotations. Such an intensity effect was observed only in single-crystal films of yttrium and bismuth iron garnets, iron, nickel, hematite and yttrium orthoferrite [38] and did not exceed 0.1% in magnitude. Both effects in polar and longitudinal polarizations are odd not only in respect to the magnetization but also in respect to the angle of deviation of the plane of polarization from the p- and s- polarizations. Similar geometries with the light of 45°-polarization allow to obtain dichroic effects and distinguish the domains with opposite magnetization orientations [18].

At the same time, quadratic in magnetization effects might appear for certain configurations. The orientation effect arising from different refractive indices for linearly polarized light in the two orthogonal in-plane configurations is quite small; however, it is observed in various ferromagnets [39,40]. A larger effect, so-called longitudinal magnetophotonic intensity effect, was recently observed in nanostructured iron-garnets due to the excitation of the guided modes and transformation of mode polarization in the in-plane magnetized medium [25,41].

Features of the distribution of p and s components inside individual plates (layers) and multilayer structures at large angles of incidence close to the Brewster angle lead to a number of new interesting effects [42–44] since the structure transparency and Faraday rotation for the p- and s-polarized light might differ several times. An ability to switch the transmitted light on and off under the application of the out-of-plane and in-plane fields was theoretically predicted in [42] for the magnetophotonic crystal supporting bound states in the continuum. Although the modulation was quite high in this case, it is rather challenging to implement the configuration in practice.

We report the odd magneto-optical linear dichroism arising in the magnetophotonic crystal (MPC) in the Faraday configuration under the oblique incidence of light with arbitrary linear (p + s) polarization. The effect manifests itself in the dependence of light transmittance with arbitrary linear (p + s) polarization in a magnetophotonic crystal (MPC) on the magnetization and the state of polarization. The effective absorption of MPC becomes a linear function of the magnetization, that allows the light modulation during magnetization reversal. The effect originates due to the different propagation and localization of p- and s-polarized light components inside the MPC and the resulting dependence of $Q$-factor of the cavity mode of MPC on the state of arbitrary linear (p + s) polarization.

## 2. Materials and Methods

### 2.1. Materials

An all-garnet MPC with magnetic cavity layer and general formula [MA / N]$^6$ /2 MA / [N / MA]$^6$ was used as object of research. Therein N and MA denote the garnets of compositions $Sm_3Ga_5O_{12}$ and $Bi_{2.97}Er_{0.03}Al_{0.5}Ga_{0.5}O_{12}$, respectively [45–46]. The former garnet layer is diamagnetic while the latter is ferrimagnetic [47]. The MPC structure was synthesized by rf-magnetron sputtering on substrate of $Gd_3Ga_5O_{12}$ (GGG) with (111) crystallographic orientation [43,45,46]. The thicknesses of MPC layers are $h_{MA}$ = 74.5 nm and $h_N$ = 99.6 nm for MA and N, respectively. The MPC photonic band gap center $\lambda_0$ and resonant wavelength $\lambda_R$ for the cavity mode have been matched to 775 nm at normal incidence.

### 2.2. Simulation

Optical and magneto-optical spectra of MPC were obtained by numerically solving Maxwell's equations by generalized matrix method 4×4 [48]. In the simulations, we take into account the initial state of light polarization, characterized by the angle $\Psi_0$ reckoned from the position of p-polarization, angle of incidence $\theta$ and the contribution of reflection from the backside of a transparent substrate. The layers of MPC were described by the components of permittivity tensors determined from the optical and magneto-optical spectra of previously synthesized single-layer films and MPCs of similar compositions. In general, the permittivity tensor of magneto-optical layers of iron garnets in optical frequency range for considered geometries has the form

$$\hat{\varepsilon}_M = \begin{pmatrix} \varepsilon_{xx} & -\varepsilon_{xy} & 0 \\ \varepsilon_{xy} & \varepsilon_{xx} & -\varepsilon_{yz} \\ 0 & \varepsilon_{yz} & \varepsilon_{xx} \end{pmatrix}. \qquad (1)$$

In the linear in magnetization approximation, the diagonal components are equal to each other, while the non-diagonal ones are determined by the medium gyration $g$: $\varepsilon_{xy} = i \cdot g \cdot \cos(\gamma)$ and $\varepsilon_{yz} = i \cdot g \cdot \sin(\gamma)$, respectively, where $\gamma$ is an angle between the $z$ axis and magnetization direction. The values of the tensor components of MA-layer at resonant wavelengths of MPC are $\varepsilon_{xx} = 6.858 + 0.014 \cdot i$, $g = -0.021 + 0.0002 \cdot i$ at $\theta = 60°$ ($\lambda_R = 721$ nm) and $\varepsilon_{xx} = 6.682 + 0.010 \cdot i$, $g = -0.018 + 0.0002 \cdot i$ at $\theta = 0°$ ($\lambda_R = 775$ nm). The permittivities of N-layer and GGG-substrate are well described by a scalar: $\varepsilon_N = 3.865$, $\varepsilon_{GGG} = 3.948$ at $\lambda_R = 721$ nm and $\varepsilon_N = 3.844$, $\varepsilon_{GGG} = 3.904$ at $\lambda_R = 775$ nm.

*2.3. Magneto-Optical Measurements*

Experimental measurements of odd magneto-optical linear dichroism were performed using the automated magneto-optical setup. The light of a halogen lamp was decomposed into a spectrum by the Czerny-Turner monochromator. Then the monochrome light passed through the optical fiber, the motorized polarizer, the first system of focusing lenses, the sample, and the second system of focusing lenses. The second system of lenses focused light on the working area of silicon photodetector, the data from which were read by an automated system and transmitted to a personal computer. A motorized polarizer had the ability to rotate through arbitrary angles to fix the polarization state of incident light. The sample was located in the gap of the electromagnet on a special holder at an angle of 60° to the angle of light incidence. For each value of the input polarization, the two spectral dependences of transmittance were recorded, in a positive $T(+M_{z_i})$ and negative $T(-M_{z_i})$ saturating magnetic field (3.5 kOe) with respect to light propagation direction. The value of odd magneto-optical linear dichroism was estimated according to the theory presented in the next section.

To measure the values of Faraday rotation angles for wavelength of cavity mode, a compensation method was used. The method is based on determining the minimum intensity of light transmitted through the MPC during rotation of the analyzer. Faraday rotation angles were determined as half-difference of two values $\Phi(+M_{z_i})$ and $\Phi(-M_{z_i})$ obtained for two opposite directions of magnetic field of the same strength, respectively:

$$\Phi = \frac{\Phi(+M_{z_i}) - \Phi(-M_{z_i})}{2}. \qquad (2)$$

Based on the measured Faraday rotation values, the $Q$-factor of cavity mode in the experiment was determined.

**3. Results**

*3.1. Model of odd magneto-optical linear dichroism and light intensity modulation in the Faraday configuration*

The experimental configuration to observe the effect is shown in Figure 1. The light with an arbitrary linear (p + s) polarization, characterized by an angle $\Psi_0$ between the polarization and plane of light incidence, is incident at an angle $\theta$ on MPC surface and, as a result of refractions and reflections from the interfaces, passes through the MPC at the wavelength of the cavity mode. In this case, the oblique incidence of light leads to a difference in the Fresnel coefficients of s- ($\Psi = 90°$) and p- ($\Psi = 0°$) polarizations. As a consequence, at oblique incidence the $Q$-factor and the transmittance of the cavity mode of MPC depend on the state of arbitrary linear (p + s) polarization. The presence of magnetization, which has a non-zero projection $M_{z_i}$ on the direction of light $z_i$, leads to the appearance of the magneto-optical Faraday rotation $\Phi$. Since any rotation of an arbitrary linear (p + s) polarization state takes the light to another linear (p + s) state with a different $Q$-factor, the transmittance of MPC will change in the presence of magnetization and will be generally a function of two parameters $T(\Psi, M_{z_i})$.

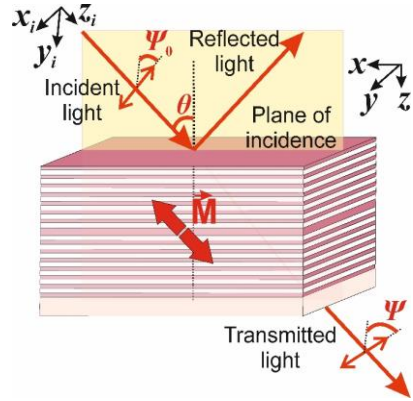

**Figure 1.** Configuration for observation of the odd magneto-optical linear dichroism.

The effect can be qualitatively interpreted using the effective length of light propagation. A multiple interference leads to the increase of the effective length inside MPC. Additionally, due to the violation of spatial symmetry, the effective length inside MPC depends on its polarization at oblique incidence. Approximately, we can assume that the effective length $L(\Psi)$ takes the following value

$$L(\Psi) = Q(\Psi) \cdot L_0, \tag{3}$$

where $Q(\Psi)$ is the quality factor of cavity mode, showing how many times the light is reflected and passed inside MPC, $L_0$ is the real total thickness of the magneto-optical layers in MPC. The quality factor depends on the state of polarization $\Psi$, since the Fresnel coefficients depend on $\Psi$.

Then, we can express the enhancement of the Faraday effect in MPC as

$$\Phi(\Psi) = \phi \cdot L(\Psi) = \phi \cdot Q(\Psi) \cdot L_0, \tag{4}$$

where $\Phi(\Psi)$ is Faraday rotation angle of cavity mode in MPC, $\phi$ is specific Faraday rotation of garnet magneto-optical layer MA.

Transmittance depends on the passed distance according to the Beer–Lambert–Bouguer law:

$$T \propto e^{-\alpha \cdot L}, \tag{5}$$

where the absorption coefficient $\alpha = 2 \cdot n'' \cdot k_0$ does not depend on the magnetization $M$ for linear polarization. However, since the polarization state is transformed due to the Faraday effect in the process of light propagation, changes in the effective length and, as a consequence, in the transmittance occur. In this case, the polarization state can be represented as a function of the coordinate $z_i$ along light incidence:

$$\Psi(z_i) = \Psi_0 + \Phi(z_i) \approx \Psi_0 + \phi \cdot Q(\Psi_0) \cdot z_i, \tag{6}$$

where $\Psi_0$ is initial state of polarization at $z_i = 0$ at the interface between MPC and air. The last approximate equality is valid if $\Phi$ is small. Here we expanded $\Phi(z_i)$ into a series in $z_i$ and took into account only the linear term in $z_i$.

Let's consider the exponent in (5) and its change over an infinitely small segment of the effective optical path $d\tilde{z}_i$:

$$-\alpha \cdot d\tilde{z}_i = -2 \cdot k'' \cdot d\tilde{z}_i = -2 \cdot k'' \cdot Q(\Psi) \cdot dz_i = -2 \cdot k'' \cdot Q_0 \cdot \left(1 + Q'_{\Psi_0} \cdot \phi \cdot z_i\right) \cdot dz_i, \quad (7)$$

$$Q_0 = Q(\Psi_0), \quad (8)$$

$$Q'_{\Psi_0} = \left.\frac{\partial Q}{\partial \Psi}\right|_{\Psi=\Psi_0}. \quad (9)$$

Summing up over the thickness of the all magneto-optical layers passed by the light, we obtain an expression for the transmittance that takes into account the changes introduced by the Faraday rotation:

$$T \propto e^{-2 \cdot k'' \cdot Q_0 \cdot \int_0^{L_0}\left(1 + Q'_{\Psi_0} \cdot \phi \cdot z_i\right)dz_i} = e^{-2 \cdot k'' \cdot Q_0 \cdot L_0 \left(1 + Q'_{\Psi_0} \cdot \phi \cdot \frac{L_0}{2}\right)}. \quad (10)$$

Thus, since the specific Faraday rotation angle of the medium $\phi$ depends on the magnetization $M_{z_i}$, the transmittance will depend on $M_{z_i}$. In this case, the effective absorption of MPC for any arbitrary linear (p + s) polarization will depend linearly on the magnetization:

$$\alpha = 2 \cdot k'' \cdot Q_0 \cdot \left(1 + Q'_{\Psi_0} \cdot \phi \cdot \frac{L_0}{2}\right). \quad (11)$$

Let's consider a few special cases. (1) Absence of magnetization, $M_{z_i} = 0$ and $\phi = 0$:

$$T \propto e^{-2 \cdot k'' \cdot Q_0 \cdot L_0}. \quad (12)$$

(2) Incident wave with polarization $\Psi_0 = 0°$ or $\Psi_0 = 90°$: from symmetry considerations we get $Q'_{\Psi_0} = 0$ that indicates the absence of an effect linear in $M_{z_i}$. (3) According to (11), it is possible to select and estimate the magnetic contribution to effective absorption if we consider the propagation of two arbitrary linear (p + s) polarizations that experience the Faraday rotation in opposite directions. It is possible to implement by magnetization reversal $+M_{z_i} \rightarrow -M_{z_i}$:

$$\Delta \alpha = \alpha\left[\Psi\left(+M_{z_i}\right)\right] - \alpha\left[\Psi\left(-M_{z_i}\right)\right] \approx -2 \cdot k'' \cdot Q_0 \cdot Q'_{\Psi_0} \cdot \phi \cdot L_0. \quad (13)$$

In the latter case, the experimentally measured difference in transmittance for arbitrary linear (p + s) polarization $\Psi_0$ will take the form

$$T(+M_z) - T(-M_z) \propto T_0 \cdot \left(\frac{e^{-a\phi} - e^{a\phi}}{2}\right) \approx$$
$$\approx -T_0 \cdot \left(k'' \cdot Q_0 \cdot Q'_{\Psi_0} \cdot L_0^2\right) \cdot \phi, \quad (14)$$

where $T(+M_{z_i})$ and $T(-M_{z_i})$ are the transmittance coefficients for the two opposite projections of the magnetization vector on the direction of light $z_i$, $T_0 = 2 \cdot e^{-2 \cdot k'' \cdot Q_0 \cdot L_0}$, $a = k'' \cdot Q_0 \cdot L_0^2 \cdot Q'_{\Psi_0}$.

Expression (14) shows that in the approximation the transmittance difference depends linearly on $M_{z_i}$ for arbitrary linear (p + s) polarization state $\Psi_0$ upon magnetization reversal as well. Thus, by analogy with the odd transverse Kerr effect, we should be able to observe the light intensity modulation when switching the magnetization in opposite directions as a result of the manifestation of odd magneto-optical linear dichroism:

$$\delta_T = \frac{2 \cdot \left(T(+M_{z_i}) - T(-M_{z_i})\right)}{T(+M_{z_i}) + T(-M_{z_i})} \cdot 100\% \tag{15}$$

$$\delta_T \propto \left[-2 \cdot k'' \cdot Q_0 \cdot Q'_{\Psi_0} \cdot L_0^2 \cdot \phi\right] \cdot 100\% = \left[\Delta\alpha \cdot L_0\right] \cdot 100\%. \tag{16}$$

*3.2. Simulation and experimental observation of the odd magneto-optical linear dichroism*

To demonstrate the effect, we chose the maximum angle of incidence among all those implemented in the experimental setup, which provides the maximum difference between the *Q*-factors of p- and s-polarized states, $\theta = 60°$. Figure 2a shows the calculated transmittance spectra of considered all-garnet MPC for various polarizations of the incident light and wavelength at an oblique incidence. The cavity mode is observed in the middle of the photonic bandgap at $\lambda_R = 721$ nm as the narrow transmittance peak for any polarization of light. At the same time, the transmittance at this resonance significantly depends on the polarization and gradually changes 4 times from $T \sim 82\%$ for $\Psi_0 = 0°$ to $T \sim 17\%$ for $\Psi_0 = 90°$ (Figure 2b, c).

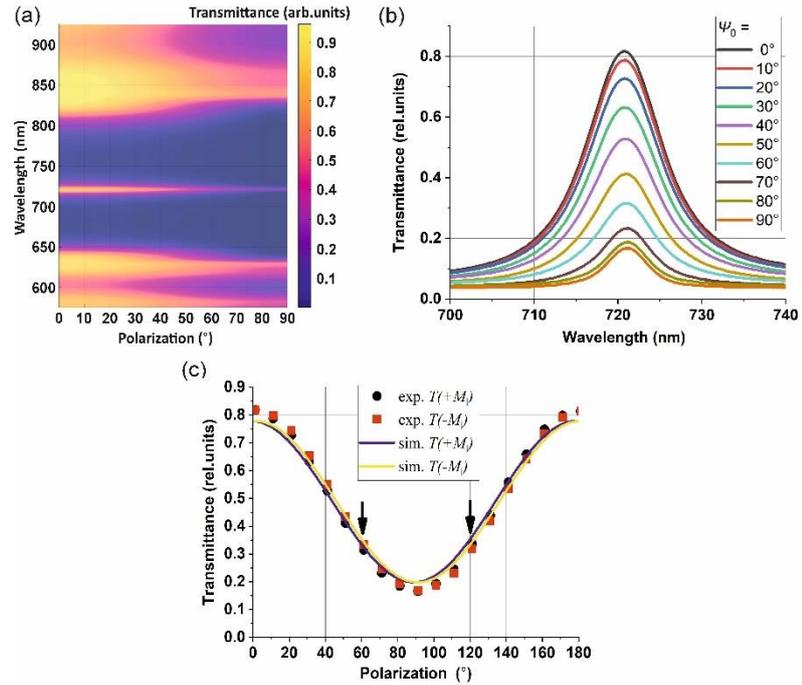

**Figure 2.** A color plot for the transmittance calculated for different wavelengths and polarization angles $\Psi$ of the arbitrary linearly (s + p) polarized light (a). Measured transmittance of MPC with positive orientation of magnetization $T(+M_{z_i})$ versus initial state of polarization $\Psi_0$ (b). Measured (symbol) and simulated (lines) transmittance of MPC at resonance wavelength $\lambda_R = 721$ nm in configurations with opposite orientations of magnetization $T(+M_{z_i})$ and $T(-M_{z_i})$ versus initial state of polarization $\Psi_0$ (c). The arrows indicate the values of $\Psi_0$ corresponding to the configurations in Figure 4c.

As it was noted above, such a feature arises due to the difference between the *Q*-factors of cavity mode for p- and s-polarized light caused by the difference of the Fresnel coefficients. S-polarization is trapped and localized inside the MPC more strongly than p-polarization, as is seen from the optical field distribution along MPC (Figure 3a). This leads to the fact that the values of the Faraday rotation angle of the MPC Φ will be different for p- and s-polarized light, that, in accordance with formula (4), allows us to estimate the behavior of the dependence $Q(\Psi_0)$ (Figure 3b).

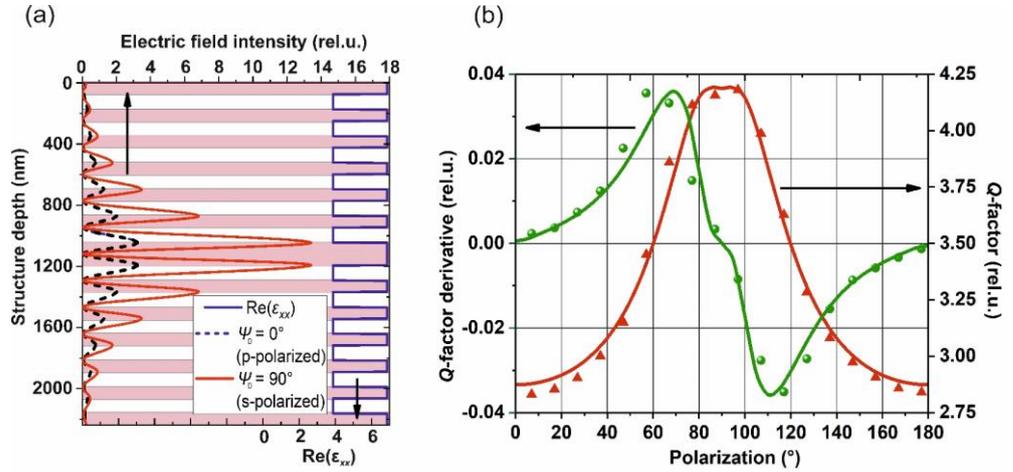

**Figure 3.** Simulated distribution of electric field intensity of p- and s-polarized light inside MPC (a) and measured (symbol) and simulated (lines) dependences of the $Q$-factor and its derivative $Q'_{\Psi_0}$ on the initial state of polarization $\Psi_0$ at $\lambda_R$ = 721 nm (b). The angle of incidence is $\theta$ = 60°.

When the polarization state changes from p to s, the Faraday rotation angle Φ and $Q$-factor of the cavity mode increases by 1.5 times. Since for the arbitrary linearly (s + p) polarized light clockwise and counterclockwise magneto-optical rotations turn the light to the states with the different $Q$-factor and transmittance, in accordance with (16), the polarization dependence of the effect of magneto-optical linear dichroism is determined mainly by the multiplication of $Q_0 \cdot Q'_{\Psi_0}$. This leads to the fact that the maximum effect is observed for the polarization states, in which $Q'_{\Psi_0}$ has a maximum (Figure 3b and 4a). Therefore, the effect reaches 6% for polarizations of $\Psi_0$ = 60° and $\Psi_0$ = 120° and vanishes for $\Psi_0$ = 0° and $\Psi_0$ = 90° ($Q'_{\Psi_0} = 0$).

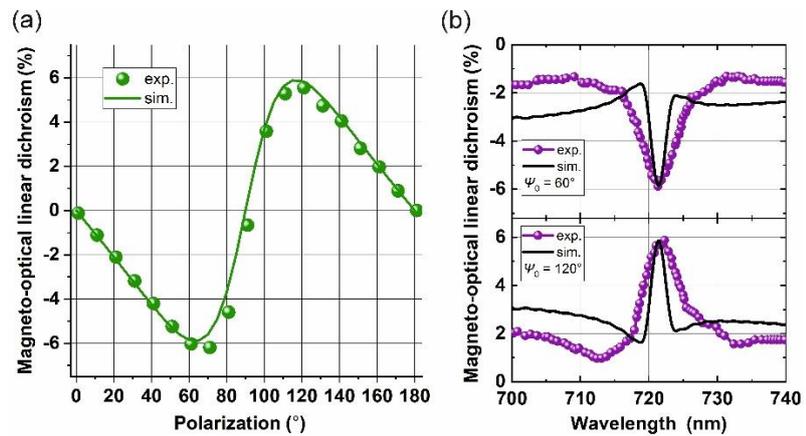

**Figure 4.** Measured (symbols) and simulated (lines) values of the odd magneto-optical linear dichroism $\delta_T$ of MPC: (a) at resonance wavelength $\lambda_R$ = 721 nm versus initial state of polarization $\Psi_0$, (b) in the vicinity of resonance for different $\Psi_0$: 60° and 120°. The angle of incidence is $\theta$ = 60°.

For $\Psi_0$ < 90° positive magnetization rotates polarization towards s-polarization, while negative magnetization rotates towards p-polarization. Therefore,

$Q\left[\Psi\left(+M_{z_i}\right)\right] > Q\left[\Psi\left(-M_{z_i}\right)\right]$ and $T(+M_{z_i}) < T(-M_{z_i})$, see Figure 4b. For $90° < \Psi_0 < 180°$ the situation becomes vice versa, and the sign of the effect changes to positive.

Similar measurements and simulations were also carried out for the sample located normal to the incident light to ensure the absence of the discussed effect. In this case, the quality factor of the structure ceases to depend on the state of polarization and is constant, i.e. $Q_0 = 6.5$, $Q'_{\Psi_0} = 0$.

## 4. Discussion

We suppose that the odd magneto-optical linear dichroism can be applied for modulators and magneto-optical visualization. It can provide more efficient light modulation, by analogy with the effect described in [42], in the case of optimization of the configuration and MPC structure, for example, when the Faraday rotation angle of the light with polarization $\Psi = 45°$ reaches values close to 45°. In this case the polarization of either s- or p-polarized state can be switched, thereby implementing the maximum possible modulation without the use of an analyzer in the observation system. In the magneto-optical microscopy the effects of linear dichroism can be efficient in the situations where the magnetic circular dichroism vanishes, for example, for the visualization of antiferromagnetic domains [18].

## 5. Conclusions

The odd magneto-optical linear dichroism in the Faraday configuration arises in a MPC at oblique incidence of the light with an arbitrary linear (s + p) polarization. The oblique incidence leads to polarization-dependent Q-factor of the cavity mode and effective absorption of the MPC, that simultaneously depends on the state of polarization and magnetization. The effect reveals itself as intensity modulation of the transmitted light with an arbitrary linear (s + p) polarization upon magnetization reversal in two opposite directions. In multiple pass mode the magneto-optical Faraday effect switches the initial state of light polarization with one Q-factor to the state with another Q-factor. It was shown that the increment of Q-factor of the cavity mode determines the character of the polarization dependence of the odd magneto-optical linear dichroism. Therefore, the maximum magnitude of the effect reaches 6% in MPC for the polarizations $\Psi = 60°$ (or $\Psi = 120°$) at the angle of incidence $\theta = 60°$. However, the effect is absent in configurations where the increment of Q-factor is zero ($\Psi = 0°$, $\Psi = 90°$ or $\theta = 0°$).

**Author Contributions:** Conceptualization, T.M. and D.I.; methodology, T.M. and D.I.; software, T.M. and S.L.; validation, D.I.; formal analysis, T.M.; investigation, S.L.; resources, V.B.; writing—original draft preparation, T.M.; writing—review and editing, T.M., D.I., V.B. and V.B.; visualization, T.M. and S.L.; supervision, V.B.; project administration, V.B.; funding acquisition, V.B. All authors have read and agreed to the published version of the manuscript.

**Funding:** This work was financially supported by the Ministry of Science and Higher Education of the Russian Federation [Megagrant project No. 075-15-2022-1108] and partially by the MSU Program of Development, project No 23-SCH06-03.

**Institutional Review Board Statement:** Not applicable.

**Informed Consent Statement:** Not applicable.

**Data Availability Statement:** The data presented in this study are available on request from the corresponding author.

**Conflicts of Interest:** The authors declare no conflict of interest.

**References**

1. Zvezdin, A.K.; Kotov, V.A. *Modern Magnetooptics and Magnetooptical Materials*; Publisher: IOP Publishing, Bristol, Philadelphia, 1997.

2. *Magnetophotonics: From Theory to Applications*; Inoue, M.; Levy, M.; Baryshev, A.; Publisher: Springer-Verlag, Berlin, Heidelberg, 2013.
3. Hu, S.; Guo, Z.; Dong, L.; Deng, F.; Jiang, H.; Chen, H. Enhanced Magneto-Optical Effect in Heterostructures Composed of Epsilon-Near-Zero Materials and Truncated Photonic Crystals. *Front. Mater.* **2022**, 9, 843265. https://doi.org/10.3389/fmats.2022.843265.
4. Chernov, A.I.; Kozhaev, M.A; Ignatyeva, D.O.; Beginin, E.N.; Sadovnikov, A.V.; Voronov, A.A.; Karki, D.; Levy, M.; Belotelov, V.I. All-Dielectric Nanophotonics Enables Tunable Excitation of the Exchange Spin Waves. *Nano Lett.* **2020**, 20(7), 5259-5266. https://doi.org/10.1021/acs.nanolett.0c01528.
5. Diaz-Valencia, B.F.; Mejía-Salazar, J.R.; Oliveira Osvaldo, N.; Jr., Porras-Montenegro, N.; Albella, P. Enhanced Transverse Magneto-Optical Kerr Effect in Magnetoplasmonic Crystals for the Design of Highly Sensitive Plasmonic (Bio)sensing Platforms. *ACS Omega* **2017**, 2, 7682–7685. https://doi.org/10.1021/acsomega.7b01458.
6. Díaz-Valencia, B.F.; Moncada-Villa, E.; Gómez Faustino, R.; Porras-Montenegro, N.; Mejía-Salazar, J.R. Bulk Plasmon Polariton Modes in Hyperbolic Metamaterials for Giant Enhancement of the Transverse Magneto-Optical Kerr Effect Molecules. *Molecules* **2022**, 27, 5312. https://doi.org/10.3390/molecules27165312.
7. Grunin, A.A.; Mukha, I.R.; Chetvertukhin, A.V.; Fedyanin, A.A. Refractive index sensor based on magnetoplasmonic crystals. *JMMM* **2016**, 415, 72–76. https://doi.org/10.1016/j.jmmm.2016.03.069.
8. Wang, Q.; Yao, H.; Feng, Y.; Deng, X.; Yang, B.; Xiong, D.; He, M.; Zhang, W. Surface plasmon resonances boost the transverse magneto-optical Kerr effect in a CoFeB slab covered by a subwavelength gold grating for highly sensitive detectors. *Opt. Express* **2021**, 29, 10546–10555. https://doi.org/10.1364/OE.414749.
9. Ignatyeva, D.O.; Kapralov, P.O.; Knyazev, G.A.; Sekatskii, S.K.; Dietler, G.; Nur-E-Alam, M.; Vasiliev, M.; Alameh, K.; Belotelov, V.I. High-Q surface modes in photonic crystal/iron garnet film heterostructures for sensor applications. *JETP Lett.* **2017**, 104, 679–684. https://doi.org/10.1134/S0021364016220094.
10. Rizal, C.; Pisana, S.; Hrvoic, I. Improved Magneto-Optic Surface Plasmon Resonance Biosensors. *Photonics* **2018**, 5, 15. https://doi.org/10.3390/photonics5030015.
11. David, S.; Polonschii, C.; Luculescu, C.; Gheorghiu, M.; Gáspár, S.; Gheorghiu, E. Magneto-plasmonic biosensor with enhanced analytical response and stability. *Biosensors and Bioelectronics* **2015**, 63, 525–532. https://doi.org/10.1016/j.bios.2014.08.004.
12. Wu, J.; Qing, Y.M. Near-Perfect TMOKE in Photonic Crystal Structures for Sensing Devices With High Figure of Merit. *IEEE Sensors Journal* **2022**, 22, 19177–19182. https://doi.org/10.1109/JSEN.2022.3204678.
13. Borovkova, O.V.; Ignatyeva, D.O.; Sekatskii, S.K.; Karabchevsky, A.; Belotelov, V.I. High-Q surface electromagnetic wave resonance excitation in magneto-photonic crystals for super-sensitive detection of weak light absorption in near-IR. *Photonics Research* **2020**, 8(1), 57-63. https://doi.org/10.1364/PRJ.8.000057.
14. Rizal, C.; Manera, M.G.; Ignatyeva, D.O.; Mejía-Salazar, J.R.; Rella, R.; Belotelov, V.I.; Pineider, F.; Maccaferri, N. Magnetophotonics for sensing and magnetometry toward industrial applications. *J. Appl. Phys.* **2021**, 130, 230901. https://doi.org/10.1063/5.0072884.
15. Ignatyeva, D.O.; Knyazev, G.A.; Kalish, A.N.; Chernov, A.I.; Belotelov, V.I. Vector magneto-optical magnetometer based on resonant all-dielectric gratings with highly anisotropic iron garnet films. *Journal of Physics D: Applied Physics* **2021**, 54, 295001. https://doi.org/10.1088/1361-6463/abfb1c.
16. Ignatyeva, D.O.; Krichevsky, D.M.; Belotelov, V.I.; Royer, F.; Dash, S.; Levy, M. All-dielectric magneto-photonic metasurfaces. *J. Appl. Phys.* **2022**, 132, 100902. https://doi.org/10.1063/5.0097607.
17. Belyaev, V.K.; Rodionova, V.V.; Grunin, A.A.; Inoue, M.; Fedyanin, A.A. Magnetic field sensor based on magnetoplasmonic crystal. *Sci. Rep.* **2020**, 10, 7133. https://doi.org/10.1038/s41598-020-63535-1.
18. Schafer, R.; Oppeneer, P.M.; Ognev, A.V.; Samardak, A.S.; Soldatov, I.V. Analyzer-free, intensity-based, wide-field magneto-optical microscopy. *Appl. Phys. Rev.* **2021**, 8, 031402. https://doi.org/10.1063/5.0051599.
19. McCord, J. Progress in magnetic domain observation by advanced magneto-optical microscopy. *J. Phys. D : Appl. Phys.* **2015**, 48, 333001. https://doi.org/10.1088/0022-3727/48/33/333001.
20. Zeeman, P. Mesures relatives du phénomène de Kerr. *Leiden Commun.* **1896**, 29.
21. Wind, C.H. On the theory of magneto-optic phenomena. *Phys. Rev.* **1898**, 6, 43. https://doi.org/10.1103/PhysRevSeriesI.6.43.
22. Grunin, A.A.; Zhdanov, A.G.; Ezhov, A.A.; Ganshina, E.A.; Fedyanin, A.A. Surface-plasmon-induced enhancement of magneto-optical Kerr effect in all-nickel subwavelength nanogratings. *Appl. Phys. Lett.* **2010**, 97, 261908. https://doi.org/10.1063/1.3533260.
23. Torrado, J.F.; González-Díaz, J.B.; González, M.U.; García-Martín, A.; Armelles, G. Magneto-optical effects in interacting localized and propagating surface plasmon modes. *Opt. Express* **2010**, 18, 15635–15642. https://doi.org/10.1364/OE.18.015635.
24. Belotelov, V.I.; Kreilkamp, L.E.; Kalish, A.N.; Akimov, I.A.; Bykov, D.A.; Kasture, S.; Yallapragada, V.J.; Achanta, Venu Gopal; Grishin, A.M.; Khartsev, S.I.; Nur-E-Alam, M.; Vasiliev, M.; Doskolovich, L.L.; Yakovlev, D.R.; Alameh, K.; Zvezdin, A.K.; Bayer, M. Magnetophotonic intensity effects in hybrid metal-dielectric structures. *Phys. Rev. B* **2014**, 89, 045118. https://doi.org/10.1103/PhysRevB.89.045118.
25. Belotelov, V.I.; Bykov, D.A.; Doskolovich, L.L.; Kalish, A.N.; Kotov, V.A.; Zvezdin, A.K. Giant Magnetooptical Orientational Effect in Plasmonic Heterostructures. *Optics Letters* **2009**, 34, 398–400. https://doi.org/10.1364/OL.34.000398.
26. Halagacka, L.; Vanwolleghem, M.; Postava, K.; Dagens, B.; Pistora, J. Coupled mode enhanced giant magnetoplasmonics transverse Kerr effect. *Opt. Express* **2013**, 21, 21741–21755. https://doi.org/10.1364/OE.21.021741.

27. Belotelov, V.I.; Zvezdin, A.K. Magnetooptics and extraordinary transmission of the perforated metallic films magnetized in polar geometry. *JMMM* **2006**, 300, e260–e263. https://doi.org/10.1016/j.jmmm.2005.10.095.
28. Chai, H.; Lu, Y.; Zhang, W. Enhancement of transverse magneto-optical Kerr effects and high sensing performance in a trilayer structure with nanopore arrays. *Results in Physics* **2021**, 31, 105049. https://doi.org/10.1016/j.rinp.2021.105049.
29. Barsukova, M.G.; Musorin, A.I.; Shorokhov, A.S.; Fedyanin, A.A. Enhanced magneto-optical effects in hybrid Ni-Si metasurfaces. *APL Photon.* **2019**, 4, 016102. https://doi.org/10.1063/1.5066307.
30. Cichelero, R.; Oskuei, M.A.; Kataja, M.; Hamidi, S.M.; Herranz, G. Unexpected large transverse magneto-optic Kerr effect at quasi-normal incidence in magnetoplasmonic crystals. *JMMM* **2019**, 47, 54–58. https://doi.org/10.1016/j.jmmm.2018.12.036.
31. Carvalho, W.O.F.; Moncada-Villa, E.; Oliveira, O.N., Jr.; Mejía-Salazar, J.R. Beyond plasmonic enhancement of the transverse magneto-optical Kerr effect with low-loss high-refractive-index nanostructures. *Phys. Rev. B* **2021**, 103, 075412. https://doi.org/10.1103/PhysRevB.103.075412.
32. Maksymov, I.S.; Hutomo, J.; Kostylev, M. Transverse magneto-optical Kerr effect in subwavelength dielectric gratings. *Opt. Express* **2014**, 22, 8720–8725. https://doi.org/10.1364/OE.22.008720.
33. Royer, F.; Varghese, B.; Gamet, E.; Neveu, S.; Jourlin, Y.; Jamon, D. Enhancement of Both Faraday and Kerr Effects with an All-Dielectric Grating Based on a Magneto-Optical Nanocomposite Material. *ACS Omega* **2020**, 5, 2886–2892. https://doi.org/10.1021/acsomega.9b03728.
34. Ignatyeva, D.O.; Karki, D.; Voronov, A.A.; Kozhaev, M.A.; Krichevsky, D.M.; Chernov, A.I.; Levy, M.; Belotelov, V.I. All-dielectric magnetic metasurface for advanced light control in dual polarizations combined with high-Q resonances. *Nat. Commun.* **2020**, 11, 5487. https://doi.org/10.1038/s41467-020-19310-x.
35. Yang, W.; Liu, Q.; Wang, H.; Chen, Y.; Yang, R.; Xia, S.; Luo, Y.; Deng, L.; Qin, J.; Duan, H.; Bi, L. Observation of optical gyromagnetic properties in a magneto-plasmonic metamaterial. *Nat. Commun.* **2022**, 13, 1719. https://doi.org/10.1038/s41467-022-29452-9.
36. Voronov, A.A.; Karki, D.; Ignatyeva, D.O.; Kozhaev, M.A.; Levy, M.; Belotelov, V.I. Magneto-optics of subwavelength all-dielectric gratings. *Opt. Express* **2020**, 28, 17988–17996. https://doi.org/10.1364/OE.394722.
37. Xia, Sh.; Ignatyeva, D.O.; Liu, Q.; Wang, H.; Yang, W.; Qin, J.; Chen, Y.; Duan, H.; Luo, Y.; Novák, O.; Veis, M.; Deng, L.; Belotelov, V.I.; Bi, L. Circular displacement current induced anomalous magneto-optical effects in high index Mie resonators. *Laser & Photonics Reviews* **2022**, 16, 2200067. https://doi.org/10.1002/lpor.202200067.
38. Krinchik, G.S.; Chepurova, E.E.; Ehgamov, S.V. Magneto-optical intensity effects in ferromagnetic metals and dielectrics. *Zhurnal Ehksperimental'noj i Teoreticheskoj Fiziki* **1978**, 74, 375–378.
39. Carey, R.; Thomas, B.W.J.; Viney, I.V.F.; Weaver, G.H. Magnetic birefringence in thin ferromagnetic films, *J. Physics D: Appl. Phys.* **1968**, 1, 1679. https://doi.org/10.1088/0022-3727/1/12/314.
40. Krinchik, S.S.; Gushchin, V.S., Magnetooptical effect of change of electronic structure of a ferromagnetic metal following rotation of the magnetization vector. *JETP Lett.* **1969**, 10, 24–26.
41. Belotelov, V.I.; Kreilkamp, L.E.; Kalish, A.N.; Akimov, I.A.; Bykov, D.A.; Kasture, S.; Yallapragada, V.J.; Achanta, V.G.; Grishin, A.M.; Khartsev, S.I.; Nur-E-Alam, M.; Vasiliev, M.; Doskolovich, L.L.; Yakovlev, D.R.; Alameh, K.; Zvezdin, A.K.; Bayer, M. Magnetophotonic intensity effects in hybrid metal-dielectric structures. *Phys. Rev. B* **2014**, 89, 045118. https://doi.org/10.1103/PhysRevB.89.045118.
42. Ignatyeva, D.O.; Belotelov, V.I. Bound states in the continuum enable modulation of light intensity in the Faraday configuration. *Optics Letters* **2020**, 45, 6422–6425. https://doi.org/10.1364/OL.404159.
43. Grishin, A.M.; Khartsev, S.I. Waveguiding in All-Garnet Heteroepitaxial Magneto-Optical Photonic Crystals. *JETP Lett.* **2019**, 109, 83–86. https://doi.org/10.1134/S0021364019020012.
44. Lyubchanskii, I.L.; Dadoenkova, N.N.; Lyubchanskii, M.I.; Shapovalov, E.A.; Zabolotin, A.E.; Lee, Y.P.; Rasing, T.. Response of two-defect magnetic photonic crystals to oblique incidence of light: Effect of defect layer variation. *J. Appl. Phys.* **2006**, 100, 096110. https://doi.org/10.1063/1.2362987.
45. Grishin, A.M. Amplifying magneto-optical photonic crystal, *Appl. Phys. Lett.* **2010**, 97, 061116. https://doi.org/10.1063/1.3479910.
46. Dzibrou, D.O.; Grishin, A.M. Fitting transmission and Faraday rotation spectra of [$Bi_3Fe_5O_{12}$/$Sm_3Ga_5O_{12}$]$^m$ magneto-optical photonic crystals. *J. Appl. Phys.* **2009**, 106, 043901. https://doi.org/10.1063/1.3195067.
47. Levy, M.; Borovkova, O.V.; Sheidler, C.; Blasiola, B.; Karki, D.; Jomard, F.; Kozhaev, M.A.; Popova, E.; Keller, N.; Belotelov, V.I. Faraday rotation in iron garnet films beyond elemental substitutions. *Optica* **2019**, 6(5), 642-646.
48. Passler, N.C.; Paarmann, A. Generalized 4 × 4 matrix formalism for light propagation in anisotropic stratified media: Study of surface phonon polaritons in polar dielectric heterostructures. *J. Opt. Soc. Am. B* **2017**, 34, 2128–2139. https://doi.org/10.1364/JOSAB.34.002128.